\documentclass[final,2p,times]{elsarticle}
\usepackage{amssymb}
\usepackage{amsmath}
\input{tcilatex}
\journal{ 
	 Mechanics Research Communications and accepted on 2 February 1976. Version  with developped calculations. \qquad\qquad\qquad\qquad\qquad\qquad\qquad\qquad\qquad\qquad\qquad\qquad\qquad\qquad\qquad\qquad}
\begin{document}

	\begin{frontmatter}
	\begin{center}

		\title{{\large NOETHER THEOREM IN FLUID DYNAMICS}\\
			\bigskip
	{\bf \normalsize Mechanics Research Communications \\ vol. 3, 131--135 (1976)} \vskip 0.2cm
	{\it \normalsize (version with developed calculations)}\\ {\normalsize https://doi.org/10.1016/0093-6413(76)90002-1} }
		
			\end{center}

		\author[Henri]{\large  Henri Gouin}

		\address[Henri]{ Aix-Marseille University,  CNRS, IUSTI
			UMR 7343,
			13013 Marseille, France 
			\\

	E-mails: henri.gouin@univ-amu.fr; henri.gouin@ens-lyon.org}

		\begin{abstract}
		Invariance theorems in analytical mechanics, such as   Noether's theorem, can be adapted to continuum mechanics. For this purpose, it is useful to give a functional representation of the motion and to interpret the groups of invariance with respect to the space of reference  associated with Lagrangian variables. A convenient method of calculus   uses  the Lie derivative. For instance,  Kelvin theorems can be obtained by such a method.
		\end{abstract}
	
\end{frontmatter}
 \section{Hamilton's action of a   fluid motion}
  The fluid motion is given by a differentiable and inversible fonction $\varphi_t$ from a reference space $\mathcal D_0$ on a space $\mathcal D_t$ which is occupied by the fluid a time $t$ 
  \begin{equation*}
 	\boldsymbol{x}={\varphi}_t(\boldsymbol{X}), \quad\boldsymbol{x}\in {\mathcal D}_{t}, \quad  \boldsymbol{X}\in {\mathcal D}_{0} 
 \end{equation*}
or by a differentiable and reversible function  $$\   \boldsymbol{z}   = \boldsymbol\Phi (\boldsymbol{Z}), \quad\boldsymbol{z}\in {\mathcal W}, \quad  \boldsymbol{Z}\in {\mathcal W}_{0} $$
\begin{equation*}
 \boldsymbol{z}=  %
\left[\begin{array}{c}
	t\\
	\boldsymbol{x}
\end{array}\right] 
\in {\mathcal W}  \quad \text{and}\quad	\boldsymbol{Z}= \left[\begin{array}{c}
		t\\
		\boldsymbol{X}
	\end{array}\right]\in {\mathcal W}_{0} 
\end{equation*} 
where $\mathcal W_0$ is the \textit{reference space} and $\mathcal W$ is the \textit{space-time}. 
Let $\Sigma_t$ be the edge of $\mathcal D_t$ and ${\partial\mathcal W}$ the edge of $\mathcal W$\,\ [1,2].\\
  For   any isentropic   motion of a fluid, we can define   the Hamilton action   
\begin{equation*}
	a = \int_{\mathcal W} \rho\,\left(\frac{1}{2}{\boldsymbol v}^\star{\boldsymbol v}-\alpha- \Omega\right) dw_{\boldsymbol z},
\end{equation*}
where ${\boldsymbol v}$ is the velocity, subscript $^\star$ denotes the transposition, the specific internal energy $\alpha$  is a function of the density $\rho$ and specific entropy $s$, $\Omega$ is the potential of external forces and $dw_{\boldsymbol z}$ is the volume element of  space-time  ${\mathcal W}$.  Hamilton action $a$ is a functional of the motion mapping $\boldsymbol{\Phi}$.

\section{Variation of the Hamilton  action}

In order to express the Hamilton principle using the variation of
Hamilton's action,
we consider  {one-parameter family of virtual motions}  dependent on a parameter $\varepsilon$ [1-5]
\begin{equation}
	{\boldsymbol{z}}=\boldsymbol{\Psi }(\boldsymbol{Z},\varepsilon ) \quad\text{with}\quad\boldsymbol{\Psi }(%
	\boldsymbol{Z},0)=\boldsymbol{\Phi}(\boldsymbol{Z})  \label{virtual motions}
\end{equation}
The scalar
$\varepsilon \in\mathbb{R}^\star$ is a small   parameter defined in a  real  open set $\mathcal O$ containing $0$    and $\boldsymbol{\Psi} $ is a \textit{4-D} regular function in  $\mathcal W_0\times \mathcal O$. 
Then, for the one-parameter family of virtual motions,
\begin{equation*}
	a =f(\varepsilon) \  \text{and its variation}\,\ \delta a\,\  \text{is defined by}\  \delta a =f^\prime(0)\,\delta\varepsilon 
\end{equation*}
where  in case of variation, differential values are denoted $\delta$ in place of $d$.\\ 
Two possibilities can be considered to obtain the variation of action $a$.
From 
\begin{equation}
	\delta \boldsymbol z =  \frac{\partial\mathbf\Psi}{\partial \boldsymbol Z}\,  \delta \boldsymbol Z+ \frac{\partial\mathbf\Psi}{\partial \varepsilon} \, \delta\varepsilon\quad \text{with}\quad \delta\varepsilon=1  \quad {\rm at} \quad \varepsilon =0\label{variationdiff}
\end{equation}
we deduce \\

$\bullet$\quad  A first variation:    \begin{equation*}\delta \boldsymbol z=   \boldsymbol{\tilde\zeta} \,\quad {\rm when}\,\ \delta \boldsymbol Z =\boldsymbol 0
\end{equation*}                     

$\bullet$\quad  A second variation:    \begin{equation*}\delta \boldsymbol Z =\boldsymbol{\hat\zeta}  \,\quad {\rm when}\,\ \delta \boldsymbol z =\boldsymbol 0
\end{equation*}
Symbols \textit{tilde}   and \textit{hat} respectively denote the first variation and the second variation  associated with a virtual motions defined by Eq. \eqref{virtual motions}.
\\
\begin{remark} 
	The two variations  are dependent.
	In fact   \eqref{virtual motions} and \eqref{variationdiff}  imply 
	$$\displaystyle
	\frac{\partial  \mathbf\Psi(\boldsymbol Z,0)}{\partial \varepsilon}+\frac{\partial  \mathbf\Psi(\boldsymbol Z,0)}{\partial \boldsymbol Z}\, \boldsymbol{\hat\zeta} =\boldsymbol 0
	$$
	which can be written 
\begin{equation}
	\boldsymbol{\tilde\zeta} +\frac{\partial \boldsymbol z}{\partial \boldsymbol Z}\,   \boldsymbol{\hat\zeta}=\boldsymbol 0\label{relvirtualdsipl},
\end{equation} 
where $\displaystyle \frac{\partial \boldsymbol z}{\partial \boldsymbol Z}$ denotes the Jacobien of transformation $\boldsymbol \Phi$.         	
\end{remark}

\subsection{Connexion between variations $\tilde\delta$ and $\hat\delta$ \cite{Bretherton}}

A variation of a tensorial quantity $\sigma$ defined on ${{\mathcal W}}$ can be written $\sigma(\boldsymbol z) =  \sigma\big(\boldsymbol{\Psi }(\boldsymbol{Z},\varepsilon )\big) $; it follows   
\begin{equation*}
	\boldsymbol{\tilde\delta}\sigma=\frac{\partial\sigma}{\partial\boldsymbol z}\,\frac{\partial\boldsymbol{\Psi }(\boldsymbol{Z},\varepsilon )}{\partial\varepsilon}
\end{equation*}
and
\begin{equation*}
	\boldsymbol{\hat\delta}\sigma=\frac{\partial\sigma}{\partial\boldsymbol z}\,\frac{\partial\boldsymbol{\Psi }(\boldsymbol{Z},0 )}{\partial\boldsymbol Z
	}\,\boldsymbol{\hat\zeta}+\frac{\partial\sigma}{\partial\boldsymbol z}\,\frac{\partial\boldsymbol{\Psi }(\boldsymbol{Z},0 )}{\partial\varepsilon}=\frac{\partial\sigma}{\partial\boldsymbol z}\,\frac{\partial\boldsymbol{\Psi }(\boldsymbol{Z},0 )}{\partial\boldsymbol Z
	}\,\boldsymbol{\hat\zeta}+\hat\delta \sigma
\end{equation*}
From Eq. \eqref{relvirtualdsipl} we obtain,
\begin{equation}
	\boldsymbol{\hat\delta}\sigma =\boldsymbol{\tilde\delta}\sigma - \frac{\partial \sigma}{\partial \boldsymbol{z}}\,	\boldsymbol{\tilde\zeta}\label{key}
\end{equation}
For the special {\it displacement} field $	\boldsymbol{\tilde\zeta} =\left[\begin{array}{c}
	0\\
	\tilde\delta \boldsymbol{x}
\end{array}\right]$, we get   \\

For the specific entropy of conservative motions, $s=s_0(\boldsymbol X)$ and consequently\quad $\tilde\delta s =0$.

The potential of external forces verifies $\Omega =\Omega(t,\boldsymbol x)$, and consequently  $\tilde\delta\Omega=(\partial\Omega/\partial\boldsymbol x)\, \tilde\delta {\boldsymbol x}$.

The material derivative $d/dt$ commutes with $\tilde\delta$ and consequently the variation of the velocity verifies $\displaystyle\tilde\delta {\boldsymbol v}  = {d(\tilde\delta \boldsymbol{x})}/{dt}$.

From $\rho\, {\rm det\,} \boldsymbol F = \rho_0(\boldsymbol X)$, where $\boldsymbol F =\partial\boldsymbol x/\partial\boldsymbol X$ is the jacobian of $\varphi_t$, we deduce 
\begin{equation*} 
\tilde\delta \rho\, ({\rm det\,} \boldsymbol F)= \rho\, ({\rm det\,} \boldsymbol F)\,
{\rm Tr}\left(\boldsymbol F^{-1}\tilde\delta F\right)=0
\end{equation*}
and consequently, 
\begin{equation*}
	\tilde\delta \rho= -\,\rho\, {\rm Tr}\left(\partial\tilde\delta \boldsymbol x/\partial  \boldsymbol x\right)=-\,\rho\,{\rm div} \,\tilde\delta\boldsymbol x
\end{equation*}
 where Tr denotes the trace operator.
From Eq. \eqref{key}, we obtain  

\begin{equation}
	\left\{
	\begin{array}{l}
		\displaystyle 
		\tilde\delta  s  = 0 \\ \\
		\displaystyle \tilde\delta \rho  =-\,\rho\, \text{div}\,\tilde \delta \boldsymbol{x} \\ \\
		\displaystyle\tilde\delta {\boldsymbol v}  =\frac{d(\tilde\delta \boldsymbol{x})}{dt}\\  \\
		\displaystyle\tilde\delta \Omega =\frac{\partial\Omega}{\partial\boldsymbol{x}}\,\tilde\delta \boldsymbol{x} 
	\end{array}\right.\qquad 
	\Longrightarrow\qquad\left\{
	\begin{array}{l}
		\displaystyle 
		\hat\delta  s  = -\frac{\partial s}{\partial\boldsymbol{x}}\, \tilde\delta \boldsymbol{x} \\ \\
		\displaystyle \hat\delta \rho  =-\, \text{div}\, (\rho\,\tilde\delta\boldsymbol{x}) \\ \\
		\displaystyle\hat\delta{\boldsymbol v}  =\frac{d(\tilde\delta \boldsymbol{x})}{dt}-
		\frac{\partial{\boldsymbol v}}{\partial\boldsymbol{x}}\, \tilde\delta \boldsymbol{x}\\  \\
		\displaystyle\hat\delta \Omega =0
	\end{array}\right.\label{table}
\end{equation}\\

Then, we obtain the variation $\delta a$ as (\footnote{For  vectors $\boldsymbol{a}$ and $\boldsymbol{b}$,   $\boldsymbol{a}%
	^{\star }\boldsymbol{b}$  is the scalar product (line vector
	$\boldsymbol{a}^{\star}$
	is multiplied by column vector $\boldsymbol{b}$); for the sake of simplicity, we also denote $\boldsymbol{a}%
	^{\star }\boldsymbol{a}= \boldsymbol{a}^2$.\\ Tensor $\boldsymbol{a} {%
		\ }\boldsymbol{b}^{\star}$ (or $\boldsymbol{a}\otimes \boldsymbol{b}$) is
	the product of column vector $\boldsymbol{a}$ by line vector $\boldsymbol{b}%
	^{\star}$.\\ Tensor $\boldsymbol{%
		1}$ is the identity, $\text{grad}$ and $\text{div}$ are the gradient and divergence operators.  }) 
\begin{equation*}
	\delta a= \iiiint_{\mathcal W} \left\{\left(\frac{1}{2}{\boldsymbol v}^\star{\boldsymbol v}-\alpha- \Omega\right)\, \hat\delta\rho+\rho\,\hat\delta\left(\frac{1}{2}{\boldsymbol v}^\star{\boldsymbol v}-\alpha- \Omega\right)\right\} dw_{\boldsymbol z}
\end{equation*} 
and by taking \eqref{table} into account
we obtain 
\begin{eqnarray*}
	\delta a &=& \iiiint_{\mathcal W} \left\{\ \left(-\frac{1}{2}{\boldsymbol v}^2+\alpha+ \rho\,\frac{\partial\alpha}{\partial\rho} + \Omega\right)\, \text{div}\, (\rho\,\tilde\delta\boldsymbol{x})\right.\\ &+&\left.\rho\, \left({\boldsymbol v}^\star\frac{d(\tilde\delta \boldsymbol{x})}{dt}-{\boldsymbol v}^\star
	\frac{\partial{\boldsymbol v}}{\partial\boldsymbol{x}}\, \tilde\delta \boldsymbol{x}+\frac{\partial\alpha}{\partial s}\frac{\partial s}{\partial \boldsymbol{x}}{\tilde\delta \boldsymbol x}  \right) \ \right\} dw_{\boldsymbol z}\\\\
	&=&\iiiint_{\mathcal W} \left\{\ \left(-\frac{1}{2}{\boldsymbol v}^2 +h + \Omega\right)\, \text{div}\, (\rho\,\tilde\delta\boldsymbol{x})\right.\\ 
	&+&\left. \frac{\partial }{\partial t}\left(\rho\, {\boldsymbol v}^\star \boldsymbol{\tilde\delta x} \right)- \frac{\partial }{\partial t}\left(\rho\, {\boldsymbol v}^\star  \right)\boldsymbol{\tilde\delta x}+\text{Tr}\left(\rho {\boldsymbol v} {\boldsymbol v}^\star \frac{\partial\tilde\delta\boldsymbol{x}}{\partial\boldsymbol{x}}\right)+\rho\left(T\,\frac{\partial s}{\partial x}-{\boldsymbol v}^\star\frac{\partial {\boldsymbol v}}{\partial {\boldsymbol x}}\right)\, \tilde\delta\boldsymbol{x}\ \right\} dw_{\boldsymbol z}\\
	&=&\iiiint_{\mathcal W}\left\{\ \text{div}\left(\rho\,\left({\boldsymbol  v} {\boldsymbol v}^\star+\left(h+\Omega-\frac{1}{2}\, {\boldsymbol v}^2 \right)\boldsymbol 1\right)\ \right)\,\tilde\delta{\boldsymbol x}+\frac{\partial }{\partial t}\left(\rho\, {\boldsymbol v}^\star \boldsymbol{\tilde\delta x} \right)\right. \\
	&+&\left. \left (-\rho\, \text{grad}^\star (h+\Omega)+ \rho\, T\,\text{grad}^\star  s- \frac{\partial (\rho{\boldsymbol v}^\star)}{\partial t}-\text{div}\left(\,\rho\,{\boldsymbol v} {\boldsymbol v}^\star\right)\right)\,\tilde\delta{\boldsymbol x} \ \right\}\, dw_z
\end{eqnarray*}
\begin{eqnarray}
	\delta a&=&  \left[\iiint_{{\mathcal D }_t}\rho\,{\boldsymbol v}^\star\tilde\delta \boldsymbol{x}\, dw_x\right]_{t_1}^{t_2} +\int_{t_1}^{t_2} \left\{\iint_{\Sigma_t}\rho \,\boldsymbol{n}^\star \left({\boldsymbol v}  {\boldsymbol v}^\star+\left(h+\Omega- \frac{1}{2}{\boldsymbol v}^2\right)\boldsymbol 1 \right)\,\tilde\delta \boldsymbol{x}\,d\sigma \right\}dt\notag\\
	&+&\iiiint_{\mathcal W }  \tilde\delta \boldsymbol{x}^\star\left\{-\rho\, \text{grad}\,(h+\Omega)+ \rho\, T\,\text{grad}\, s- \frac{\partial (\rho{\boldsymbol v}^\star)}{\partial t}-\text{div}\,\rho\,{\boldsymbol v} {\boldsymbol v}^\star  \right\}\,dw_z \label{deltaa} 
\end{eqnarray}
where   $\boldsymbol{n}$ is the unit normal vector external to $\Sigma_t$, $\displaystyle h=\alpha+ \frac{\partial\alpha}{\partial\rho}$ is the enthapy, $\displaystyle T=\frac{\partial\alpha}{\partial s}$ the Kelvin temperature, $dw_x$ the volume element of ${\mathcal D }_t$, $d\sigma $ the volume element of $\Sigma_t$, and $\displaystyle [\ \ ]_{t_1}^{t_2}$ denotes the difference of values between times ${t_2}$ and ${t_1}$. 
\subsection{Field $\tilde\delta\boldsymbol{x}$ leaving the Hamiltonian invariant}
\medskip
For any motion, the vector field $\tilde\delta\boldsymbol{x}$ such that   $\delta a =0$ must verify  
\begin{equation*}
	\hat\delta\, s =0, \quad\hat\delta\, \rho =0,\quad \hat \delta\, {\boldsymbol v} =\boldsymbol{0} 
\end{equation*}
or  from Eqs \eqref{table}, these connditions are equivalent to 
\begin{equation}
	\frac{\partial s}{\partial\boldsymbol{x}}\, \tilde\delta\boldsymbol{x} =0, \quad \text{div}\, (\rho\,\tilde\delta \boldsymbol{x})  =0,\quad  \frac{d(\tilde\delta \boldsymbol{x})}{dt}-
	\frac{\partial{\boldsymbol v}}{\partial\boldsymbol{x}}\, \tilde\delta\boldsymbol{x} =\boldsymbol{0} \label{key2}
\end{equation}
Relation \eqref{key2}$^3$ means that $\tilde\delta\boldsymbol{x}$ is a vector field of ${\mathcal D }_t$ with a zero Lie's derivative with respect to the velocity field $\boldsymbol{v}$; then $\displaystyle\frac{\partial s}{\partial\boldsymbol{x}}\, \tilde\delta\boldsymbol{x}$ and  $\text{div}\, (\rho\,\tilde\delta \boldsymbol{x})$  have a zero Lie's derivative with respect to the velocity field $\boldsymbol{v}$ (see \cite{Lie-gouin} for properties), and
{\it theorem 15} in \cite{Lie-gouin} allows to write
\begin{equation*}
	\rho\,\tilde\delta \boldsymbol{x} = f(s,\eta)\,\ \text{grad}\, s\times\text{grad}\, \eta\quad  \text{with}\quad \frac{d\eta}{dt} =0
\end{equation*}
When $\tilde\delta \boldsymbol{x}$ is null on $\Sigma_t$ the second integral of Eq. \eqref{deltaa} is null.  Due to the equation of motion expressed in thermodynamic form, when we choose the real motion (corresponding to $\varepsilon =0$), the third integral of  Eq. \eqref{deltaa} is also null. 
Therefore, according to Hamilton's principle which requires the nullity of the variation of Hamilton's action [1], we obtain 
\begin{equation}
	\left[\iiint_{{\mathcal D }_t}\rho\,{\boldsymbol v}^\star\tilde\delta \boldsymbol{x}\, dw_x\right]_{t_1}^{t_2}
	=0\label{discont}
\end{equation}
\section{Case of perfect fluids}
Consider an isentropic closed fluid curve $(\gamma_0)$ in reference space $\mathcal D_0$ determined by the intersection of a surface $\sigma(\boldsymbol X) =  \sigma_0$, where $\sigma$ is a differentiable function and $\sigma_0$ a constant, and a surface $s(\boldsymbol X)=s_0$, where $s_0$ is a constant.\\
The position of   image  $(\gamma_t)$ of $(\gamma_0)$ in $\mathcal D_t$ is determined by the intersection of $s(\varphi_t(\boldsymbol x) )=s_0$ and $\sigma(\varphi_t(\boldsymbol x)) =\sigma_0$ and thus $\displaystyle \frac{d\sigma}{dt} =0$.\\
The  curve $(\gamma_t)$ being closed, we can choose $\sigma$ so that the four surfaces
\begin{equation*}
	\sigma=\sigma_1,\,\  \sigma=\sigma_2,\,s=s_1,\,\  s=s_2, \quad \text{where}\quad \sigma_1<\sigma_0<\sigma_2\,\ \text{and}\,\ s_1<s_0<s_2
\end{equation*}
delineate a domain containing $(\gamma_t)$.\\
\begin{figure}[h]
	\begin{center}
		\includegraphics[scale=0.8]{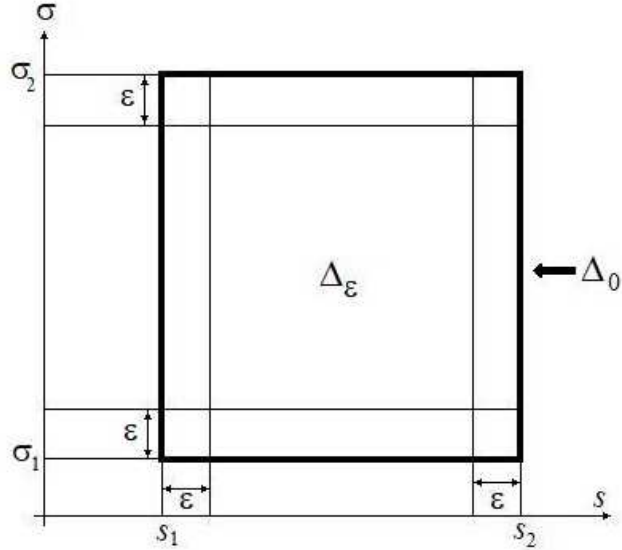}
	\end{center}
	\caption{domain $\Delta_0$ and domain $\Delta_\varepsilon$}
	\label{Fig.1}
\end{figure}
Let us consider the domain $\Delta_0$, set of points $(s,\sigma)$ such that 
\begin{equation*}
 s_1\leq s_0\leq s_2 \quad\text{and} \quad 	\sigma_1 \leq \sigma_0 \leq\sigma_2
\end{equation*}
and domain $\Delta_\varepsilon$ set of points $(s,\sigma)$ such that 
\begin{equation*}
s_1+\varepsilon \leq s \leq s_2-\varepsilon\,\ \text{and}\,\  \sigma_1+\varepsilon \leq \sigma \leq\sigma_2-\varepsilon \end{equation*} 
with 
\begin{equation} 
	\varepsilon <\frac{1}{2}\,\text{inf}\,(\sigma_2-\sigma_1, s_2-s_1) \label{ineq}
\end{equation} 
For  $\varepsilon$ verifying \eqref {ineq},  by using Tietze-Urysohn's theorem, it is possible to obtain a positive application  $(s,\sigma)\longrightarrow g(s,\sigma)$ majored by $1$, of value $1$ inside the set $\Delta_\varepsilon$ and being null on the complementary of $\Delta_0$.\\ Let consider the domains $\mathcal D_{t,1}$ image of $\Delta_0$  and 
$\mathcal D_{t,1,\varepsilon}$ image of
$\Delta_\varepsilon$  by the application $\boldsymbol\varphi_t$ at $t$ value. The vector field 
\begin{equation*}
	\rho\,\tilde\delta \boldsymbol{x}=k\,g(s,\sigma)\,\text{grad}\, s\times\text{grad}\, \sigma
\end{equation*}
where $k$ is  a constant, leaves invariant the Hamilton action and is zero on $S_{t,1}$. For such a field we have relation \eqref{discont}. Then, $\rho, \boldsymbol v, \tilde\delta \boldsymbol{x}$ being continous fonctions of $\boldsymbol{x}$ and $\mathcal D_{t,1}$ being a compact set, 
\begin{equation*}
	\forall  \,\eta\in\mathbb{R}^\star, \ \exists\,\varepsilon\in\mathbb{R}^\star\ \text{verifying} \ \eqref{ineq} 
	\end{equation*}
such that,
	\begin{equation*}
	\left|\iiint_{{\mathcal D }_{t_i,1}}\rho\,{\boldsymbol v}^\star\tilde\delta \boldsymbol{x}\, dw_x- \iiint_{{\mathcal D }_{t_i,1,\varepsilon}}\rho\,{\boldsymbol v}^\star\tilde\delta \boldsymbol{x}\, dw_x\right| 
	<\eta
\end{equation*}
where $i =1, 2$. Then,  in this inequality, we use in \eqref{discont} the vector field $\rho\, \tilde\delta \boldsymbol{x}$ defined by 
\begin{equation}
\rho\, \tilde\delta \boldsymbol{x} = \frac{1}{(s_2-s_1)\,(\sigma_2-\sigma_1)}\, \text{grad}\, s\times\text{grad}\, \sigma\label{condition}
\end{equation}
and when $\varepsilon$ goes to zero, we can replace $\Delta_\varepsilon$ by $\Delta_0$.\\
We get as curvilinear  coordinates $\sigma, s, \ell$ where $d\ell$ is the element of length of curves $(\gamma_{t,{s_c},{\sigma_c}})$ intersections of   surfaces $s(\boldsymbol x)=s_c$ and $\sigma(\boldsymbol x)=\sigma_c$, where $s_c$ and $\sigma_c$ are two constants.\\
Let us consider  the unit vector field $\boldsymbol \tau$, tangent to $(\gamma_{t,{s_c},{\sigma_c}})$ in the 
direction of  $\text{grad}\, s\,\times\,\text{grad}\, \sigma$, we deduce the volume element  
\begin{equation*}
	dw_{\boldsymbol x}=\frac{\text{grad}\, s\times\text{grad}\, \sigma}{( \text{grad}\, s\times\text{grad}\, \sigma)^2}\, \boldsymbol \tau\ d\ell\,  ds\, d\sigma
\end{equation*}
 and by using expression \eqref{condition}, we obtain
\begin{equation*}
	\forall\, i \in \{1, 2\},\quad \iiint_{{\mathcal D }_{t_i,1}} \rho\,\boldsymbol{v}^\star \tilde\delta\boldsymbol{x} dw_{\boldsymbol x}=\iint_{\Delta_0}\left(\int_{\gamma_{t_i,{s},{\sigma}}}\boldsymbol{v}^\star\boldsymbol{\tau}\, d\ell\ \right) \frac{ds\,d\sigma}{(s_2-s_1)\,(\sigma_2-\sigma_1)}
\end{equation*}
After having multiplied by $(s_2-s_1)\,(\sigma_2-\sigma_1)$, when $s_1$ and $s_2$ go to $s_0$ and $\sigma_1$ and $\sigma_2$ go to $\sigma_0$,  we deduce  from Eq. \eqref{discont}     
\begin{equation*}
	\int_{\gamma_{t_1}}\boldsymbol{v}^\star\boldsymbol{\tau}\, d\ell=\int_{\gamma_{t_2}}\boldsymbol{v}^\star\boldsymbol{\tau}\, d\ell
\end{equation*}
Noting that $t_1$ and $t_2$ have no special role, we get
\begin{equation*}
	\frac{d}{dt}\left(\int_{\Gamma_{t}}\boldsymbol{v}^\star\boldsymbol{\tau}\, d\ell\right)=0
\end{equation*}
Consequently, we obtain the two theorems:

\begin{theorem} {\bf Kelvin 1:}
	For a perfect fluid, the circulation on a closed and isentropic fluid curve moving with the flow is constant in the motion.
\end{theorem}

\begin{theorem} {\bf Kelvin 2:}
	A fluid being barotropic, for any fluid curve moving with the flow, the velocity circulation is conserved in the motion.
	\end{theorem}
We have on integral form, a first integral representing thermodynamic invariants. This result comes from the choice of $\tilde\delta \boldsymbol{x} $.\\

{\it Consequences :}\\

The research of these first integrals has been made from the reference space $\mathcal W_0$ on which  relations \eqref{key2} are simply expressed.
The family of considered virtual displacements $\tilde\delta\boldsymbol{x}$ corresponding to the vector field $\boldsymbol{\tilde\zeta} =\left[\begin{array}{c}
	0\\
	\tilde\delta \boldsymbol{x}
\end{array}\right]$ is indeed interpreted as a permutation of the particles in the space of motions which have the same density, the same specific entropy and the same velocity field, and which are localized by the Lagrangian variables on the reference space. Such a permutation makes Hamilton's action invariant.

We did not consider the vector field $\left[\begin{array}{c}
	\tilde\delta t\\
	\boldsymbol{0}
\end{array}\right]$; it is easy to see that this field allows to obtain the writing of the entropy equation.\\

These first integrals correspond to the most general motions of perfect fluids. Particular motions and particular thermodynamic laws can give less general first integrals corresponding to fields from a family of virtual motions making the action always invariant. 

\section{Conclusion}
Particles described in the Lagrangian variables with the same density, entropy and velocity field can be interchanged in the Eulerian variables.
Such a permutation of particles does not modify the Hamilton action  
and the first integrals obtained are forms of Kelvin's theorem.

	\end{document}